\documentclass[12pt,letterpaper]{JHEP3}
\usepackage{epsfig}

\def\I1{{\mathbb 1}}
\def\lsim{\mathrel{\rlap{\lower 4pt \hbox{\hskip 1pt $\sim$}}\raise 1pt \hbox
    {$<$}}}
\def\gsim{\mathrel{\rlap{\lower 4pt \hbox{\hskip 1pt $\sim$}}\raise 1pt \hbox
    {$>$}}}

\def \da9{D9-\bar{D}9}

\def \n{\noindent}

\def \beq{\begin{equation}}
\def \eeq{\end{equation}}
\def \da9{D9-\bar{D}9}
\def \dap{Dp-\bar{D}p}

\title{Inflation from Tachyon Condensation, Large $N$ Effects}
\author{Mahbub Majumdar$^1$ \& Anne-Christine Davis$^2$ \\

$^1$ Theoretical Physics Group\\ Imperial College, Blackett Lab\\ Prince
Consort Road, London SW7 2BW, UK\\
$^2$ Department of Applied Mathematics and Theoretical Physics, \\
   Centre for Mathematical Sciences\\ University of Cambridge,
   Wilberforce Road, Cambridge, CB3 0WA, U.K.\\
Emails: {\tt m.majumdar@ic.ac.uk, A.C.Davis@damtp.cam.ac.uk}

}

\abstract{Using only general properties of the tachyon potential we
show that inflation may be generic when many branes and anti-branes
become coincident. Inflation may occur because of: (1) the assistance of
the many diagonal tachyon fields; (2) when the tachyons condense in a
staggered fashion; or (3) when some of them condense very late.  We point
out that such inflation is in some sense a stringy implementation of
chaotic inflation and may have important applications for
``regularizing'' a lopsided or singular cosmological compact surface.

 }

\date{\today}

\begin{document}


\section{Introduction}

Inflation corresponds to an instability -- a period of accelerated
growth of the universe.  Field theoretic tachyons likewise correspond
to instabilities and similarly arise whenever string theoretic
instabilities appear. For example, branes oriented at generic angles
are unstable and possess tachyons~\cite{polchinski}.  Systems with
non-coincident $Dp$ and $D(p-2)$ branes are unstable and possess
tachyons~\cite{narain}.  Also, non-BPS branes and brane anti-brane
pairs are unstable and are tachyonic~\cite{sen}.  Such tachyons
generate the dynamics of the systems but, eventually disappear once
the systems have flowed to a stable and supersymmetric state.  The
correspondence between tachyons and rapid time dependent behaviour is
suggestive of some cosmological importance and the challenge posed by
this paper is to link inflationary instabilities with tachyonic
ones.  Some previous attempts are~\cite{rollcosmo}.  We find that configurations with $N$ coincident $\dap$ branes
and anti-branes may generate inflation if $N\gg1$.

The early universe was very hot and thus a stringy understanding of
cosmological time-dependence will require a study of string
instabilities at high temperature.  As discussed in ~\cite{danielsson}
, there exists evidence, at least in the case of non-BPS branes and
brane anti-brane systems, that tachyons disappear at high
temperature.  Essentially, tachyons in such a system
correspond to negative curvature of a ``tachyon'' potential.  At high
temperature, the curvature becomes positive as symmetry of the
potential is restored in the usual thermal field theory sense.  Hence,
the tachyons disappear.  However, once the universe expands, the
temperature falls (if the expansion is adiabatic).  At a critical
temperature $T_c$, the tachyon potential changes shape and tachyons
appear.

As shown in~\cite{paper3} these tachyons can lead to the production of D-brane
remnants and any leftover high dimensional defects need to be diluted
away.  We will show that tachyons which condense late may dilute away
problematic defects.

The central result of this paper is: inflation may be generic in
non-supersymmetric string theories.  Whenever, a patch of the
universe is covered with many coincident branes and anti-branes or
non-BPS branes, tachyons appear and inflation may result.  In some
sense tachyonic inflation is a stringy implementation of chaotic
inflation.

This kind of inflation may not generate appropriate density
perturbations.  However, the goal of Planckian inflation shouldn't
necessarily be the solution of all of cosmology's problems.  It would
be extraordinary if one quantum gravity mechanism at $10^{19}$GeV were
responsible for structure formation at a few eV via the generation of
acceptable density perturbations.  A more realistic aim of Planckian
inflation is the prevention of early collapse of the universe, i.e. a
solution of the flatness problem.

Later bouts of inflation can generate acceptable density perturbations
and cure the ills prior episodes of inflation did not, or other
mechanisms such as a curvaton mechanism~\cite{curvaton} can be used to
generate perturbations. There exists some prejudice that inflation
should occur only once.  However, we take the view that if inflation
can happen once, it can probably happen many times during the history
of the universe. No inflation after the first bout means that
$\ddot{a} \le 0$ forever.  It is difficult to believe that any
quantity which can be positive or negative will forever choose one
sign over the other.  Strong no-go theorems would be required and none
exist.  It is true that normal matter obeying the strong energy
condition does not give $\ddot{a} > 0$.  However, the universe may
contain all sorts of matter, e.g. topological defects, vacuum energy,
etc.  In fact, the universe now seems to be filled with some sort of
quintessential vacuum energy causing it to accelerate
today~\cite{cmbexp}.

This paper is organized as follows.  We review some generalities about
the tachyon potential from the vacuum string field theory point of
view in section \ref{tac}. We show how the potential changes for a
large number of branes in section \ref{N}.  Then we discuss
inflationary scenarios in section \ref{inf}.  There we describe what
happens when: (1) all the tachyons condense at the same time and show
that this is an example of assisted inflation as in~\cite{anupum};
(2) what happens when they condense in a staggered fashion, which we
call staggered inflation; (3) finally, what happens when some tachyons
condense very late.  Next, we briefly discuss density perturbations in
section \ref{den}.  Finally, we end with several comments, criticisms
and make the analogy between chaotic and tachyonic inflation.

\section{Tachyon Potential}\label{tac}

String theory possesses many extra scalar and gauge fields. These
fields are not fixed, and their dynamics are uncertain because their
potentials are unknown. Even when a non-flat potential is written down
-- it is never more than an inspired guess of the non-perturbative
physics.  Tachyon physics is rather different, as the form of the
potential is known up to two derivative terms. This is rather
striking, as the tachyon is an intrinsically stringy field and its
potential is {\em background independent}.  Recall, that the potential
takes the uniform form, where $t$ is the tachyon field and $\tau$ is  time,

\beq
V(t) = 2 \tau_p (1 + v(t))
\eeq

\n where $v(t)$ is a universal function and is the same for D-branes
wrapped on cycles of an internal compact manifold, D-branes in the
presence of a background metric, or anti-symmetric tensor field, etc.
Only the multiplicative factor, $\tau_p$ (the brane tension) is model
dependent.

The tachyon potential truncated at level (2,4) for a non-BPS brane
system is~\cite{tachyonreview}

\beq
V(t) = \tau_p(1-0.87t^2 + 0.21t^4)
\label{sftpot}
\eeq

\n This is similar to the potential of a brane anti-brane system this
is the form of the potential which we shall use. The potential is
characterized by the presence of a global minimum at a finite distance
from $t=0$ in field space, and its double well shape.

The above potential is not completely sufficient to describe the time
dependence of the tachyon field.  Once the tachyon starts to roll down
from $t=0$, it will couple to a countably infinite number of other
string fields, complicating the analysis of the time dependence.  In
fact, an appropriate treatment of time dependence would entail
knowledge of an infinite number of time derivatives of the string
field.  This can seen by noting that the Witten star product $ *$,
and its generalizations, e.g. $\hat{*}$, are related to the
non-local Moyal product~\cite{sliver}, $*_M$, which acting between two functions
$f$ and $g$ acts as

\beq
f *_M g(y) = \exp\left ( \frac{i}{2} \theta^{\mu \nu}
\frac{\partial}{ \partial x^{\mu} } \frac{\partial}{ \partial y^{\nu}
} \right ) f(x) g(y) |_{x=y}
\eeq

\n where the $\theta_{\mu \nu}$ are constants. Thus, the Moyal product
involves an infinite number of derivatives, and an infinite number of
time derivatives for $\theta_{0\mu} \neq 0$.  Hence, by analogy the
string field theory action, which involves the Witten star product,
also involves an infinite number of time derivatives.  For example, in
the bosonic case, every three point vertex is accompanied by a factor
$\exp( -\ln(4/3\sqrt{3}) \alpha'\partial^2)$, and thus involves an
infinite number of derivatives.

Hence, a string field theory description of the time dependent tachyon
condensation will involve an {\em infinite} number of time
derivatives, and specific cosmological consequences may depend on
higher derivative corrections. Nevertheless, as our discussions will
depend only on the gross features of the tachyon potential and not on
its particular shape, we will use a potential like
\ref{sftpot}.

We will not use a DBI effective action because: (1)  the
questionable validity of the DBI action when the tachyon is very
near the top its potential -- the region for $t$ which we are most
interested in;  (2) we are interested in multiple coincident
branes and there is ambiguity on how to write the DBI effective
actions of non-Abelian theories.

\section{Tachyon Condensation on $N$ $\dap$ Brane Anti-brane Pairs}\label{N}

A system of $N$ $\dap$ brane anti-brane pairs will possess a $U(N)
\times U(N)$ gauge symmetry.  The tachyon connecting the branes to the
anti-branes transforms in the bifundamental $(N,N)$ representation of
the gauge group.  Thus, such a system generally possesses $N^2$
tachyons $t_{i\bar{j}}$ where $1 \le i,\bar{j} \le N$.

The potential for such a configuration will be similar to the potential in
(\ref{sftpot}) with $t^2$ replaced by $tt^{\dagger}$ and $t$ replaced
by an $N\times N$ matrix.  The tachyon potential can then be written
as the truncated effective potential

\beq
V(t) = 2 \tau_p -  c_1 {\rm tr}( tt^{\dagger})  + c_2  (
{\rm tr}(tt^{\dagger})^2)  + {\cal{O}}(|t_i|^6)
\eeq

\n
which mimics the shape of $V(t)$ in  (\ref{sftpot}) and possesses a minimum
at $t_0$.

Using the $U(N)\times U(N)$ gauge symmetry the tachyon can be
diagonalized at {\em any} point in time, such that for $u$ and $v$ in
the first and second $U(N)$'s of the $U(N) \times U(N)$ symmetry we
have

\beq
u t v = {\rm diag}(t_1,\ldots, t_N)
\eeq

\n where the $t_i$ are complex scalar fields. The potential
then becomes

\begin{equation}
V(t_1,..., t_N) = N\tau -  c_1 \sum_1^N |t_i|^2 +
c_2 \sum_1^N |t_i|^4  + {\cal{O}}(|t_i|^6)
\label{manypotential}
\end{equation}

At the minimum of the potential (\ref{manypotential}), $V(t_1 =
t_0,...t_N = t_0) = 0$, and the individual $t_i$ satisfy the same
equation as that satisfied by a {\em single} brane anti-brane pair
tachyon at the minimum of the potential,

\beq  2\tau_p - c_1
|t_i|^2 + c_2 |t_i|^4 + {\cal{O}}( |t_i|^6) = 0
\eeq

\n Thus the tachyons are non-interacting and behave like the tachyon of
a single $\dap$ brane anti-brane pair.  However, to diagonalize the
tachyon at {\em every} point in its evolution, would require time
dependent $u(\tau)$ and $v(\tau)$ matrices. This would not necessarily
yield a continuous trajectory for the $N$ diagonal tachyons $t_i$.
Thus, it is impossible to isolate the tachyonic degrees of freedom in
$N$ tachyonic fields which are differentiable throughout their entire
trajectories.  One can however from the outset restrict the tachyon to
an Abelian part of $U(N)\times U(N)$ by setting the off-diagonal terms
to zero. Then, although the branes and anti-branes are coincident, the
worldvolume tachyons will not interact.  Hence, $t$ will then take the
form

\beq  t = {\rm
diag}(t_1,...,t_N)
\label{tdiag}
\eeq

\n and the potential will take the form (\ref{manypotential}) and the
kinetic term will also be diagonal.

A generic tachyon matrix will possess off-diagonal terms, and hence
the restriction is somewhat unphysical.  However, if $N \gg 1 $ large diagonal blocks may exist in the $N$ by $N$
tachyon matrix.


In \cite{rolling} the existence of a two parameter set of solutions of a
single time dependent (rolling) tachyon field was demonstrated.  Here,
we show that a similar $2N$ parameter set of solutions for $N$
independent tachyons exists.  We can show that time dependent solutions
with (\ref{tdiag}) exist by solving the cubic string field theory
equations of motion at level  (0,0).

At level 0, the string field is

\beq
| \Psi \rangle = c_1|0\rangle_g \otimes t|0\rangle_m
\eeq

\n where $c_1$ is a $c$ ghost,$|0\rangle_g$ and $|0\rangle_m$ are the
ghost and matter parts of the vacuum state.  The equation of motion of
$t$ is

\beq
Q_B  | \Psi \rangle = 0 ~~\Rightarrow ~~ (\Box + m^2)t = 0
~~ \Rightarrow ~~ (\Box + m^2) t_i = 0
\label{eqofmotion}
\eeq

\n where $Q_B$ is the BRST operator. Equation (\ref{eqofmotion}) has a
solution of

\beq
\tilde{t}_i = A_i e^{\tau} + B_i e^{-\tau}
\eeq

\n for spatially homogeneous tachyons.  Here $\tau$ is the physical
time; $A_i,B_i$ are set by the initial conditions of the tachyons
$t_i(0)$ and $dt_i/d\tau|_0$. As in \cite{rolling}, we can then
construct a two parameter family of solutions for {\em each} time dependent
tachyon labeled by the initial position of the tachyon and its initial
velocity.

Now, whenever a system possesses $N$ scalar fields, $N-1$ fields can
usually be set to zero, and only one field be allowed to roll to the
bottom of the potential.  For example, in the system given by the
action with an $O(N)$ symmetry

\beq
S = \frac{1}{2} \int d^q x \left (\sum_{i=1}^N (\partial \phi_i)^2 -
(\sum_{i=1}^N \phi_i^2 - \mu^2)^2 \right )
\eeq

\n any path of steepest descent given by $(\phi_1(t),...,\phi_n(t))$ can
be rotated by an $O(N)$ rotation into the path $(\phi_1(t),0,...,0)$.
However, for this to occur, $\phi_1$ must have have support over
$[0,\mu]$.

Similar arguments do not apply to the tachyon case because one tachyon
mode cannot lead to the annihilation of all $N$ $\dap$ brane
anti-brane pairs.  Each tachyon mode is defined only over $[0,t_0]$,
and can lead only to the annihilation of only one brane anti-brane
pair; it is not defined over the interval $[0,N t_0]$.

Thus tachyon condensation genuinely involves the rolling of $N$
independent tachyon fields.

\section{Inflationary Scenarios}\label{inf}

We now describe several scenarios where inflation occurs because
of tachyon condensation.  The amount of inflation is heavily
dependent on the manner in which the tachyons condense -- whether
(1) the all condense at once; (2) in a staggered fashion; and (3)
whether there is a group of tachyons which condense very late.

As is well known, sufficient conditions for inflation are: (1) the
energy density is potential dominated; (2) the motion is
over-damped; (3) the potential is sufficiently flat such that the
energy density is roughly constant

\begin{equation}
\left (\frac{\dot{a}}{a} \right )^2\sim  V(t)   \sim V(0);~~~~
\ddot{t} \ll 1 ~~~~  \Rightarrow   a(\tau) \sim a(0) \exp{\sqrt{V(0)} \tau}
\end{equation}

\n where $t$ is a scalar field generating inflation.  The potential
will be flat and the energy will be potential dominated if
$\ddot{t}$ is small and $V$ doesn't strongly vary with $t$,
i.e. if the slow roll parameters, $\eta, \epsilon$ are small

\begin{equation}
\eta   \equiv   m_p^2 \left | \frac{V''}{V}\right | \ll 1; ~~~~~
\epsilon  \equiv   m_p^2 \left |\frac{V'}{V}\right |^2 \ll 1
\end{equation}

If the slow roll conditions are satisfied, then it can be shown that
inflation will occur for $n$ efolds, where

\beq
n= \int_{\tau_i}^{\tau_f} Hd\tau \sim  - \int
\frac{V(t_i)}{V'(t_i)} dt_i.
\eeq

Although near $t_i \approx 0$ the tachyon potential is flat
($V'(0)=0$), an individual tachyon field will satisfy the slow roll
conditions for only a small portion of time.  Furthermore, vacuum
fluctuations will displace the field by a magnitude $H/(2\pi)$,
causing the field to start ``rolling'' at $t \sim H/(2\pi)$ and not at
$t_i=0$.  Because tachyon condensation occurs at an energy scale
$E\sim m_s $, where $m_s$ is typically the string scale, we will have
$H\sim m_s$, and the minimum of the tachyon potential will be $t_0\sim
m_s$. Hence, perturbations will typically displace the tachyon field
far enough away from $V(0)$ to make any inflation very
short-lived. Thus, it is unlikely that a single tachyon will lead to
the necessary number of efolds for inflation to be successful (about
60 efolds for 4D physics).  However, the situation may change for
$N\gg 1$ tachyon fields.

Suppose that $N$ independent tachyons exist.  Because of their
independence they may condense at different times, and will condense
whenever a perturbation dislodges one of them away from the maximum of the
tachyon potential. Alternatively, they could all condense
approximately simultaneously. These alternatives give rise to
different inflationary scenarios, which we elucidate below.

\subsection{Simultaneous Condensation: Assisted Inflation}

Suppose that the tachyons condense at roughly the same time.  Then the
tachyons will follow identical trajectories, which for convenience we
identify as $t_1$.  Then we can write

\beq
\sum_{i=1}^N V(t_i)  = N V(t_1) \equiv \tilde{V}(\tilde{t})
\eeq

\n where we have written the potential in terms of a single field
$\tilde{t}$ which is related to $t_1$ as

\beq
\tilde{t} = \sqrt{N} t_1 ~~~ \tilde{c}_1 = c_1 ~~~ \tilde{c}_2 =
\frac{c_2}{\sqrt{N}}
\eeq

The slow roll parameters for this $N$ field configuration,
$\epsilon_N$ and $\eta_N$ are then related to the slow roll parameters
of a single tachyon, $\epsilon$ and $\eta$ as

\beq
\epsilon_N = \frac{\epsilon}{N}; ~~~~ \eta_N = \frac{\eta}{N}
\eeq

This $1/N$ suppression is a well known feature of {\em assisted
inflation}~\cite{assisted}.  Thus,
for large $N$, the slow-roll parameters are shrunk dramatically, and
multi-tachyon inflation becomes feasible when single tachyon inflation
is not.

We now qualify what we mean by simultaneous condensation.  If it takes
a time $\Delta \tau$ to condense, then by ``roughly condense at the
same time,'' we mean that all the tachyons condense in a time $\kappa
\Delta \tau$ where $\kappa \sim {\cal{O}}(1)$ is a constant of order
one.  This means that if a small number of tachyons have condensed
while the remainder are still condensing that slow-roll is not
violated. In essence, slow roll is insensitive to small numbers of
tachyons (small relative to $N$) condensing before the bulk of them
have condensed.  This is because the energy density in the early
tachyons is small relative to the energy density of the rest of the
tachyons.

However, if the tachyons start condensing at widely differing times,
then as in other assisted inflation scenarios, no general attractor solution pushing the tachyons toward identical
trajectories exists.  Only the trivial attractor $t_i = t_j = t_0$
at the bottom of the potential exists.  For example, defining $\psi_i
\equiv t_i - t_1$

\begin{equation}
\ddot{t}_i + (D-1) H \dot{t}_i = \frac{\partial V}{\partial t_i}; ~~~~
\ddot{t}_j + (D-1) H \dot{t}_j = \frac{\partial V}{\partial t_j}; ~~~~
\Rightarrow \ddot{\psi}_i + (D-1) H \dot{\psi}_i = 2\frac{\partial
V}{\partial \psi _i}
\end{equation}

Because, $\psi_i(t) \neq 0$, $\forall t$, initially different tachyon
profiles will remain different until late times. Only at late times,
when the tachyons have reached the minimum of the potential $t_0$, are
the tachyons equal because,

\beq
\frac{\partial V}{\partial \psi_i} = 2c_1\psi_i + 12 c_2  t_1 \psi_i (\psi_i
+t_1) = 0 ~~~~\Rightarrow~~~~ \psi = 0~~{\rm at}~~ t_i =t_0
\eeq

In addition to vacuum fluctuations, thermal fluctuations will kick the
tachyons up and down their potentials.  In our scenario, tachyon
condensation begins at the end of a second order phase transition at
$T= T_c$.  However, tachyon profiles will not get frozen in until the
the temperature drops to the Ginzburg temperature $T_G < T_c$.  Large
thermal fluctuations may persist while $T_G < T < T_c$.  For example,
for a Mexican hat type potential like (\ref{sftpot}), $V(t) = g_s
(|t|^2 - |t_0|^2)^2$, the high temperature effective thermal potential
will be ~\cite{shellard}

\beq
V_{eff}(t, T) = m^2(T) |t|^2 + g_s |t|^4;~~~~~~ m^2(T) \sim g_s
(T^2 - T_c^2).
\label{therm}
\eeq

Minimizing the potential gives $|t|^2 \sim (T^2_c - T^2)$.  The
characteristic length of the system is $r_c \sim |1/m|$. When
fluctuations in the potential energy, $ \delta E_{pot} \sim r^d \delta
V$, where $d$ is the number of spatial dimensions, are comparable to
the thermal energy, $\delta E_{thermal} \sim T$, then for fluctuations
on a scale of $r_c$, in $3+1$ dimensions we find

\beq
r_c^3 m^2 t \delta t \sim T ~~~~ \Rightarrow \left |\frac{\delta t}{t}
\right | \sim g_s \frac{ T}{m}
\eeq

The Ginzburg temperature is defined by $g_s T = m$.  Thus for $T_G < T
< T_c$ fluctuations may be large.  Using the second half of
(\ref{therm}), one can show that the $T_G \sim (1-g_s) T_c$.  If $g_s$
is not very small, then $T_c$ and $T_G$ may be significantly
different, allowing sufficient time for large thermal fluctuations to
cause $t_i$ and $t_j$ to lose coherence.


\subsection{Staggered Condensation: Staggered Inflation}

We now analyse the case when the tachyons do not condense at the same
time. If the tachyons condense at widely differing times, they can be
thought of as condensing serially; i.e.  in a sufficiently short time
interval, $\tau^*$, at most one tachyon condenses.  This means that
the $N$ tachyons condense according to a Poisson distribution.  If the
time interval $\tau^*$ is sufficiently long such that one tachyon will
very likely have condensed in an interval, $\tau^*$, then we can say
that one tachyon will condense on average in a time interval of $\sim
\tau^*$, and all the tachyons will have condensed within a time $\sim
N\tau^*$. Then in the case of a brane anti-brane system the potential
will decrease by $2\tau_p$ in a time $\sim \tau^*$. The potential in
this case then takes the form

\beq
V(\tau) = 2\tau_p(N- \frac{\tau}{\tau^*})
\label{linearpot}
\eeq

Let us rewrite the slow roll parameters in a way more amenable to this
time dependent situation.  For a time dependent potential $V(\tau)$,

\beq
V(\tau) = V(0) + V'(0) \tau + V''(0) \frac{\tau^2}{2} +
{\cal{O}}(\tau^3)
\eeq

\n If $\tau$ is a time scale of order $\sim 1/m_p$, then the slow-roll
conditions can loosely be written as

\begin{equation}
\eta   \equiv   \frac{1}{m_p^2} \left |\frac{V''}{V} \right | \ll 1; ~~~~
\epsilon  \equiv   \frac{1}{m_p^2} \left |\frac{V'}{V} \right | \ll 1
\end{equation}

\n which for the potential (\ref{linearpot}) gives

\beq
\epsilon_N  \sim \frac{1}{N(\tau)m_p\tau^*}; ~~~~  \eta \sim 0
\eeq

Hence, for $\tau^*$ no smaller than a Planck time ($1/m_p$), there is a
$1/N(\tau)$ suppression in the slow roll parameters.  (Note, since $N$
is the number of brane anti-brane pairs left over at time $\tau$, $N$
is a function of time).  Thus, inflation would seem to naturally
occur, although because of the staggered nature of the tachyon
condensation the density perturbations may be large.

Suppose instead that the tachyons condense more frequently such that
in a time $\tau^*$ many tachyons may condense. For definiteness,
suppose that in a time step $\tau^*$ each tachyon has a probability of
$1/2$ of condensing.  Then the tachyons will condense via a binomial
distribution. For example, in the first time step half of the tachyons
will condense, in the second time step one fourth will condense,
etc. The time dependent potential will then be

\beq
V(\tau) = 2N\tau_p  2^{-\tau / \tau^*} = 2N\tau_p  \exp \left (-
\frac{\tau}{\tau^*} \ln 2 \right)
\eeq

\n which is an exponential potential. It gives rise to the slow roll
parameters

\beq
\eta_N  \sim \left | \frac{\ln 2}{m_p\tau^*}\right |; ~~~~  \epsilon \sim \left | \frac{\ln 2}{m_p\tau^*}\right |^2
\eeq

\n which is small for sufficiently large $\tau^*$.  This is similar to
power law inflation arising from an exponential potential in that the
slow roll parameters are constant.  However, unlike power law
inflation, any extant inflation will end when all the tachyons have
condensed.

\subsection{Late Condensation}

We now analyse what happens when a group of tachyons condense
late.  If $N$ is large, then in any scenario, it is likely that
there will always be some very late condensing tachyons. This can
happen as follows. Each tachyon will start condensing once it is
kicked off the top of the tachyon potential. The rms value of such
a kick by a vacuum fluctuation is $H/2\pi$. Since $H \sim m_s$,
most tachyons will receive a very large kick. If the size of this
kick is given by a Gaussian probability distribution, then for $N
\gg1$, there may be some tachyons out in the wings of the
distribution which receive very small kicks. Hence, they will
start rolling very close to the top of the potential, as opposed
to say a third of the way down as for the other tachyons.  The top
of the potential is necessarily flat $\epsilon = 0$ (because of
the need for $U(1)$ invariance). The curvature $V''$ is not zero
and hence $\eta \neq 0$ at the top. However, if the number of late
simultaneously condensing branes is small, but greater than one,
say three or four, then $\eta $ may also be suppressed. Thus, for
$N\gg 1$, a few tachyons might start to slow roll and condense
long after the other tachyons have ``finished'' condensing. This
is very interesting as such late condensing tachyons may dilute
defects left behind by earlier tachyons. In order for this to
work, these late tachyons must condense only to a tachyon gas (and
eventually radiation) and not create any defects by themselves.
This is likely, as entropic arguments in~\cite{paper3} suggest
that tachyons prefer condensing to vacuum rather than lower
dimensional $D$-branes.  Dilution will not require a great deal of
inflation, only perhaps 10-25 efolds. Such inflation by late
condensing fields would be philosophically similar to thermal
inflation whereby late rolling inflatons dilute troublesome
inhomogeneities at energies much lower than $m_{GUT}$.

\section{Density Perturbations} \label{den}

If all the tachyons condense at the same time, then the density
perturbations will be the same as of a single redefined tachyon field
$\tilde{t}$ and a Hubble constant defined by the redefined potential
$\tilde{V}(\tilde{t})$.  The curvature perturbations and spectral
index will then be

\beq
P(k) = \left (\frac{H}{2 \pi}\right) \frac{H^2}{\dot{\tilde{t}}^2};~~~~~
1-n =  6 \epsilon_N - 2\eta_N
\eeq

Isocurvature perturbations will be produced if the tachyons oscillate
transversely to their classical path.  Because, each tachyon is a
complex field and possesses a Mexican hat type potential, the
orthogonal direction is the angular direction.  Oscillatory behaviour
may thus occur if the $U(1)$ symmetry of the vacuum manifold is broken and
a mass, $m^2_{\theta}$ is given to the angular excitations. However,
the $U(1)$ is spontaneously broken and the angular tachyon mode is eaten up to
make all gauge fields very massive.  Thus no oscillatory behaviour
occurs.


Even if isocurvature perturbations are produced, they will not
contribute to the curvature perturbation.  This is because the path in
field space is not curved, i.e.

\beq
\dot{\theta}_{ij} = 0 ~~~~{\rm where} ~~~~ \tan{\theta_{ij}}= \frac{\dot{t}_j}{\dot{t}_i}
\eeq

\n and hence, the isocurvature perturbations do not source curvature
perturbations\footnote{For example consider the case of two fields
$t_1$ and $t_2$. Isocurvature perturbations, $\delta s$, orthogonal to
the classical trajectory, can source adiabatic fluctuations, $\delta
\sigma$, which are tangent to the inflaton field trajectory because
the evolution equations for adiabatic fluctuations, $\delta \sigma$,
may depend on $\delta s$.  For example~\cite{wands},

\beq
3H \dot{\delta\sigma} + \left( V_{\sigma\sigma} - \dot\theta^2
  \right) \delta\sigma = 2\left(\dot\theta\delta s\right)^. - 2
  \left({V_{\sigma}\over\dot\sigma} + {\dot{H}\over H}\right)
  \dot\theta\delta s
\eeq

\n where $V_{\sigma \sigma}$ and $V_{\sigma}$ are the first and second
derivatives of the potential in the direction tangential to the
trajectory. Most importantly, isocurvature perturbations will only act
as a source if $\dot{\theta} \neq 0$ where $\tan \theta =
\dot{t}_1/\dot{t}_2$; i.e. if the trajectory is curved.}. The path is
not curved because in the assisted case, $t_1 = t_2 =\cdots = t_N$.
Because, isocurvature perturbations are suppressed outside the horizon
and are difficult to observe, we conclude that they are not
particularly relevant in our case.


If the string scale is not much smaller than the Planck scale,
then density perturbations due to gravitational waves may be very
large~\cite{lindestuff}.  However, if all of the tachyons condense
simultaneously, then as shown by~\cite{assistedgravpert} the amplitude of
gravitational waves may  be suppressed by a factor of $1/N$.

However, if the tachyons do not simultaneously condense and late
condensing tachyons exist, the observable curvature perturbations and
spectral index may change.  The tachyons which are most important are the
last slowly rolling tachyons.  If the last tachyon to condense does so
after all the others have condensed, the perturbations will largely be
that of a lone tachyon during an era with a smaller Hubble
constant.

\section{Comments, Criticisms, Chaotic  Inflation and Conclusions}

We now list some potential criticisms and discuss  how our scheme
may provide a stringy implementation of chaotic inflation and
comment on how the large $N$ scenarios may be combined with a more
complete description of brane cosmology.

{\bf Criticisms:} Some criticisms of our schemes are that they use
large $N$ effects to obtain inflation and rely on a finite brane
tension as a vacuum energy source.

The assisted inflation version partially relies on a
non-zero brane tension.  To understand its effect, suppose that
$t_{in}$ is the initial value of the inflaton field and that there is
no initial energy ($V(t_{in})=0$).  One still obtains the $\epsilon_N =
\epsilon/N, \eta_N = \eta/N$ suppression.  However, $\epsilon$ and
$\eta$ may then be rather large:

\beq
\epsilon \equiv \frac{m_p^2}{2}\left(\frac{V'(t_{in})}{t_{in} V'(t_{in})} \right)^2 = \frac{m_p^2}{2t_{in}^2};~~~~~~~~~
\eta \equiv m_p^2\left (\frac{V''(t_{in})}{t_{in}^2 V''(t_{in})} \right ) = \frac{m_p^2}{t_{in}^2}
\eeq

The initial tachyon field value is $t_{in} \approx 0$, making
$\epsilon$ and $\eta$ diverge. This is overcome by taking $N$ to be
large and not taking $t_{in} =0$.  For example, due to vacuum
fluctuations, it is much more reasonable to take something like
$t_{in} = 1/10$ in string units.  Thus, $\epsilon, \eta
\sim 100$ which is not very large and can be suppressed without taking
$N$ to be exorbitantly large.

For the staggered case, a non-zero brane tension would mean that no
background energy existed and thus no staggered inflation would occur.
This is somewhat troublesome, because in realistic cases the brane tensions may vanish.
For example, in Type I theory, the negative energy of orientifolds may
cancel the positive brane energy.

From a philosophical point of view, using a non-dynamical vacuum
energy to generate inflation is not always prudent.  There are
many types of vacuum like the susy breaking vacuum energy,
electroweak phase transition breaking vacuum energy, etc., and one
must carefully argue why a chosen vacuum energy is relevant while
others are not. We argue that because branes are physical objects
with an energy density, their vacuum energy is much more relevant
than a vacuum energy resulting from a not yet understood
cancellation of zero point energies, etc.

Furthermore, using brane tensions to generate inflation is
analogous to a feature in virtually all inflationary models -
assuming the inflaton is miraculously displaced significantly up
the potential to create an initial vacuum energy $V(\phi_{in})$.
In this respect, brane tension inflation is similar to other
models of inflation.

Another possible criticism is our use of  the cubic string field
theory tachyon potential (\ref{sftpot}) for ease of use. Instead
we could have used the tachyon potential exact up to two
derivative terms~\cite{ger}.  The exact two derivative action
possesses a non-canonical kinetic term, making it harder to work
with. The exact tachyon potential is $e^{-T^2/4}$, where $T$ is
the tachyon field in boundary string field theory formalism. In
the case of a $U(N) \times U(N)$ symmetry, $T$ is a matrix.  If we
choose $T$ to be diagonal, such that $T =$ diag($T_1,...,T_N$)
then the potential will be $V = \exp (-\sum_i T_i^2/4) = \prod_i
\exp (-T_i^2/4)$.  This means the $T_i$ are interacting, and
assisted inflation fails whenever the component fields interact.
However, by moving to the canonically normalized case via the
field redefinition $t = $~erf$(T)$, and expanding the potential
about $T=0$, the potential takes the form~\cite{zwiebachfield}

\beq V(t={\rm erf}(T)) = a_0 -a_1 t^2 + a_2 t^4 + \cdots
\label{erf} \eeq

\n where $a_0,a_1,a_2$ are constants. This potential is of the
same form as (\ref{sftpot}), and happily there appears to be no
mixing of the the $t_i$, allowing assisted inflation to occur.

Unbroken supersymmetry is often needed by high energy inflation models
to keep the potential flat.  For example $H$ dependent terms can
appear increasing the mass of the fields.  Supersymmetry breaking is
not so disastrous for our tachyonic inflation models.  The assumptions
of large $N$ decreases the need for the individual tachyons to have
flat potentials.  Also, because of background independence the shape
of the tachyon potential is fixed.  Except for the value of the brane
tension, the potential is the same in a curved (cosmological)
background as it is in a flat background; extra terms spoiling
flatness such as $H^2$ terms do not contribute to the potential.

{\bf Chaotic Inflation:} Chaotic inflation gets around the problem
of displacing the inflaton, $\phi$, by arguing that it will take
different values in different Hubble regions.  If the number of
Hubble regions is large, then thermal interactions may cause some
regions to have a suitably displaced $\phi > m_p$, which allows
the vacuum energy $V(\phi > m_p)$ to source inflation.  Tachyonic
inflation requires branes to provide  vacuum energy to source
inflation.  As described in~\cite{kogan}, branes may be thermally
produced because of the diverging free energy of their open string
gases. They may wrap various cycles of an initial compact manifold
and are analogous to the displaced field values of the inflaton in
chaotic inflation.  Hence, if a (compact) patch of spacetime finds
itself wrapped by many thermally produced non-BPS branes or branes
and anti-branes, it will inflate just as a patch with $\phi > m_p$
will inflate in chaotic inflation.  This provides a stringy
mechanism for implementing chaotic inflation.   This is
particularly attractive if the brane vacuum energy source
eventually disappears via tachyon condensation (no stable lower
dimensional remnant branes created), or if small numbers of
tachyons condense late, diluting any remaining defects.  (The late
tachyons are themselves unlikely to condense to anything but
vacuum as shown in~\cite{paper3}).

{\bf Some Comments:} This leads to an interesting brane world
scenario.\footnote{We thank Ian Kogan and Steve Abel for related
discussions.} Various dimensional branes and anti-branes wrap an
initially compact spacetime.  Higher dimensional branes if not
coincident attract each other and find each other first as shown
in~\cite{paper4}. This leads to tachyon condensation and sometimes
inflation of the cycles they were wrapping.  The spacetime which
inflates will inevitably contain embedded $D3$ branes and
$\bar{D}3$ anti-branes which did not manage to find each other
before the higher dimensional branes in which they were embedded
annihilated and eventually inflated.  Thus after inflation a
spacetime will be created in which the predominant species of
branes are $D3$ (or $\bar{D}3$) braneworlds and $D1$ and
$\bar{D}1$ strings.  Thus, although observers on a braneworld will
see only their braneworld and none others and hence believe that
their universe is very finely tuned to produce only one
braneworld; in reality it will not finely tuned. Other branes do
exist, but are simply too far off to see\footnote{Note, in
principle because of interactions with other branes (assume no
supersymmetry) an observer on a braneworld would be be able to
feel the presence of other branes. However, it is likely that if
other branes are close enough to be felt by an observer on a
braneworld that  the density of branes is large enough to cause
the universe to collapse.  Thus, in viable scenarios, an observer
should not be able to easily detect the presence of other
branes.}.

\EPSFIGURE[l]{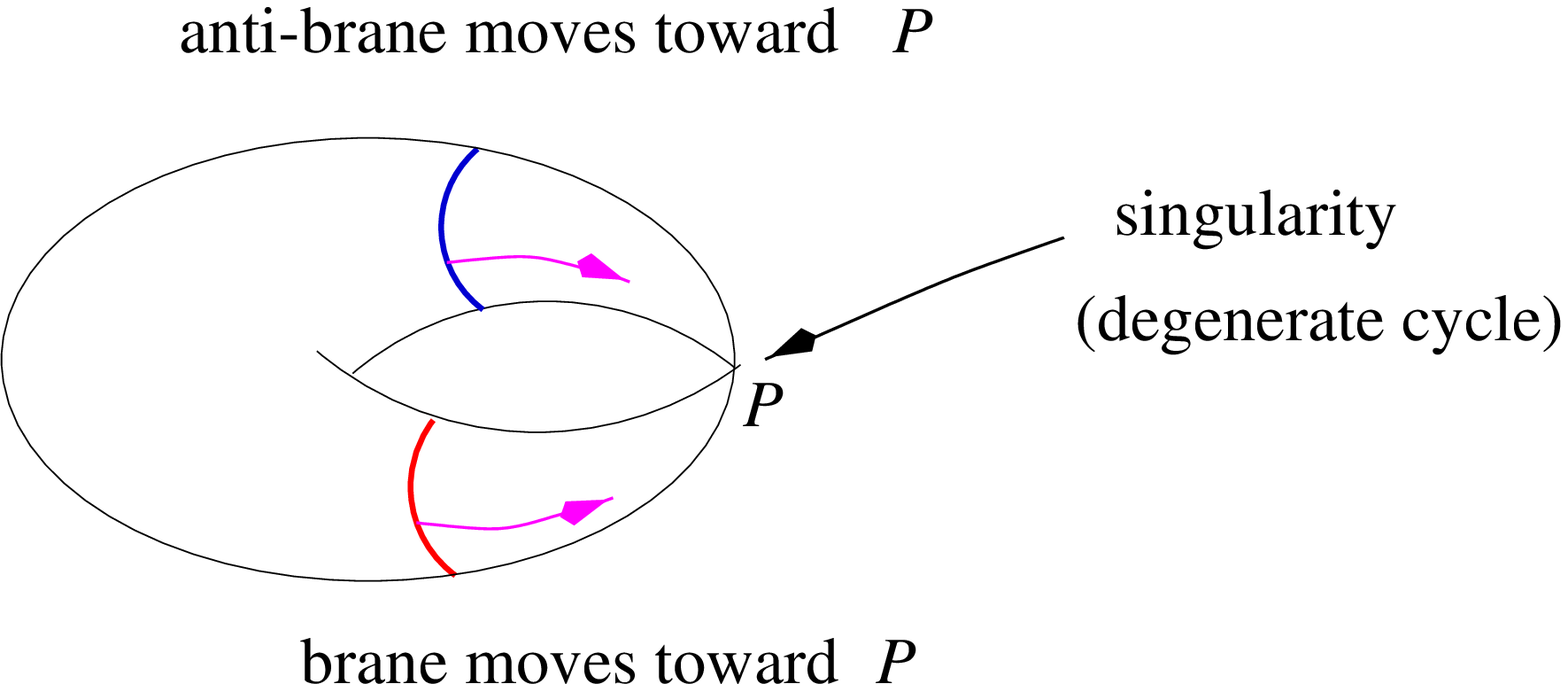,width=6cm}{Branes and
anti-branes migrate toward the zero volume cycle at $P$.  After
they reach $P$, they may inflate $P$ and resolve the singularity
there. \label{degeneratecycle}}

Finally, we reiterate the motivation for tachyonic inflation.
Although, density perturbations may be suppressed in the assisted
inflation case, even if they are excessively large tachyonic
inflation is still very interesting.  Early bouts of tachyonic
inflation may cure many cosmological problems (horizon, brane
remnant, etc) and prevent the early universe from collapsing.
Later bouts may provide the appropriate density perturbations,
etc. Inflation depends only on the sign of $\ddot{a}$.  Thus if
the sign can be positive at one instant of time during the
evolution of the universe, it is concievable it may be positive at
many instants in time.  Otherwise a selection rule must exist
explaining why $\ddot{a}$ always chooses one sign and not the
other.

We finish with one speculative comment. One situation where one
may very realistically obtain many coincident branes and
anti-branes is at points where some cycles of a compact manifold
shrink or become degenerate. Generic compact manifolds will
possess ``pinching'' singularities like the point $P$ in figure
\ref{degeneratecycle} and more generally the volume of the cycles
of the compact surface will not be constant and may vary. Branes
and anti-branes wrapping cycles of the manifold will minimize
their energy and will try to move toward 
points where the homology cycles have minimum volume. Figure
\ref{degeneratecycle} shows the case where one cycle
degenerates to a point.  Such singularities or points of minimum
volume  will thus act as a sink for branes and anti-branes and
once the branes  reach them (assuming it takes a
finite time~\cite{greene}) they will be parallel and coincident.
It is not inconceivable that these coincident brane and anti-brane
pairs may then inflate, resolving singularities like $P$ by {\em
literally} blowing them up.  This has important applications to
any compact initial state of the universe.  As generic manifolds
are singular and not regular, it is likely that the early
universe's compact manifold possessed degenerate cycles and cycles
with varying volume. Our inflationary mechanism is possibly a
means for removing singularities or points of small cycle volume,
and making the early universe more regular.

Note added: Parts of this paper were presented at the UK Cosmology
Meeting (Durham), Sept. 17, 2000, DAMTP Relativity Lunch Seminar
(Cambridge), Nov. 3, 2000; Physics of Extra Dimensions Conference
(Paris), June 25, 2001, String Phenomenology Summerfest (Sussex),
July 5, 2001, and Beyond the Standard Model Workshop (Ambleside),
August 28, 2001.

\section*{Acknowledgements}
We thank Carsten Van de Bruck, Fernando Quevedo and Neil Turok for
discussions.  A.C.D thanks PPARC. M.M also thanks the Cambridge
Commonwealth Trust, Isaac Newton Trust, DAMTP, Hughes Hall and PPARC for
financial support.



\newpage
\begin{thebibliography}{19}        



\bibitem{polchinski}
J.~Polchinski, ``Dirichlet-Branes and Ramond-Ramond Charges,'' Phys.\
Rev.\ Lett.\ {\bf 75}, 4724 (1995) [arXiv:hep-th/9510017].


\bibitem{narain}
E.~Gava, K.~S.~Narain and M.~H.~Sarmadi, ``On the bound states of p-
and (p+2)-branes,'' Nucl.\ Phys.\ B {\bf 504}, 214 (1997)
[arXiv:hep-th/9704006].


\bibitem{sen}
A.~Sen, ``Stable Non-BPS States in String Theory,'' JHEP {\bf 9806}
007 (1998), A.~Sen, ``Stable Non-BPS Bound States of BPS D-branes,''
JHEP {\bf 9808} 010 (1998), A.~Sen, ``Tachyon Condensation on the
Brane Antibrane System,'' JHEP {\bf 9808} 012 (1998), A.~Sen, ``Stable
non-BPS D-particles,'' Phys.Lett. {\bf B441} 133 (1998),
A.~Sen,``SO(32) Spinors of Type I and Other Solitons on
Brane-Antibrane Pair,'' JHEP {\bf 9809} 023 (1998), and for reviews,
see: A.~Sen, ``Non-Bps D-Branes In String Theory,'' Class.\ Quant.\
Grav.\ {\bf 17}, 1251 (2000), J.~H.~Schwarz, ``Non-BPS D-brane
systems,'' arXiv:hep-th/9908144, A.~Lerda and R.~Russo, ``Stable
non-BPS states in string theory: A pedagogical review,'' Int.\ J.\
Mod.\ Phys.\ A {\bf 15}, 771 (2000) [arXiv:hep-th/9905006].


\bibitem{rollcosmo}
D.Choudhury, D.Ghoshal, D.P.Jatkar and S.Panda,
 hep-th/0204204; 
 M.Fairbairn and M.H.Tytgat, hep-th/0204070; G.W.Gibbons,
hep-th/0204008; G.Shiu and I.Wasserman, hep-th/0205003; 
 A.Frolove, L.Kofman and A.Starobinsky, hep-th/0204187; 
 L.Kofman and A.Linde, hep-th/0205121; 
 M.C.Bento, O.Bertolami and A.A.Sen, hep-th/0208124; 
 Y.Piao, R.Cai, X.Zhang and Y.Zhang, hep-ph/0207143; 
 Y.Piao, Q.Huang, X.Zhang and Y.Zhang, hep-ph/0212219; M.Sami, P.Chingangbam and T.Qureshi, hep-th/0205179; 
 M.Sami, hep-th/0301140; H.A.Feldman and R.H.Brandenberger, Phys.Lett. {\bf B227} (1989) 359; 
 G.N.Felder, A.Frolov, L.Kofman and A.Linde, Phys.Rev. {\bf D66}
 (2002) 023507; Z.~K.~Guo, Y.~S.~Piao, R.~G. ~Cai, Y.~Z.~Zhang,
``Inflationary Attractor from Tachyonic Matter,''
[arXiv:hep-ph/0304236]


\bibitem{danielsson}
U.~H.~Danielsson, A.~Guijosa and M.~Kruczenski, ``Brane-antibrane
systems at finite temperature and the entropy of black branes,''
hep-th/0106201.

\bibitem{paper3}
M.~Majumdar and A.~C.~Davis, ``Cosmological creation of D-branes
and anti-D-branes,'' JHEP {\bf 0203}, 056 (2002)
[arXiv:hep-th/0202148].



\bibitem{curvaton}
D.~H.~Lyth, C.~Ungarelli and D.~Wands, ``The primordial density
perturbation in the curvaton scenario,'' arXiv:astro-ph/0208055.


\bibitem{cmbexp}
C.~B.~Netterfield {\it et al.}, {\it A measurement by
BOOMERANG of multiple peaks in the angular power spectrum of the
cosmic microwave background}, \astroph{0104460}; C.~Pryke {\it et
al.}, {\it Cosmological parameter extraction from the first season of
observations with DASI}, \astroph{0104490}; R.~Stompor {\it et al.},
{\it Cosmological implications of the MAXIMA-I high resolution cosmic
microwave background anisotropy measurement}, \astroph{0105062}.

\bibitem{anupum}
A.~Mazumdar, S.~Panda and A.~Perez-Lorenzana, ``Assisted inflation
via tachyon condensation,'' Nucl.\ Phys.\ B {\bf 614}, 101 (2001)
[arXiv:hep-ph/0107058].


\bibitem{tachyonreview}
K.~Ohmori,
``A review on tachyon condensation in open string field theories,''
arXiv:hep-th/0102085.

\bibitem{sliver}
M.~R.~Douglas, H.~Liu, G.~Moore and B.~Zwiebach, ``Open string star as
a continuous Moyal product,'' JHEP {\bf 0204}, 022 (2002)
[arXiv:hep-th/0202087].

\bibitem{rolling}
A.~Sen, ``Time and tachyon,'' arXiv:hep-th/0209122; A.~Sen, ``Time
evolution in open string theory,'' JHEP {\bf 0210}, 003 (2002)
[arXiv:hep-th/0207105]; A.~Sen, ``Field theory of tachyon matter,''
Mod.\ Phys.\ Lett.\ A {\bf 17}, 1797 (2002) [arXiv:hep-th/0204143];
A.~Sen, ``Tachyon matter,'' arXiv:hep-th/0203265; A.~Sen, ``Rolling
tachyon,'' JHEP {\bf 0204}, 048 (2002) [arXiv:hep-th/0203211].

\bibitem{assisted}
A.~R.~Liddle, A.~Mazumdar and F.~E.~Schunck, ``Assisted inflation,''
Phys.\ Rev.\ D {\bf 58}, 061301 (1998)
[arXiv:astro-ph/9804177];E.~J.~Copeland, A.~Mazumdar and N.~J.~Nunes,
``Generalized assisted inflation,'' Phys.\ Rev.\ D {\bf 60}, 083506
(1999) [arXiv:astro-ph/9904309]; A.~Mazumdar,
S.~Panda and A.~Perez-Lorenzana, ``Assisted inflation via tachyon
condensation,'' Nucl.\ Phys.\ B {\bf 614}, 101 (2001)
[arXiv:hep-ph/0107058]; P.~Kanti and K.~A.~Olive, ``On the realization
of assisted inflation,'' Phys.\ Rev.\ D {\bf 60}, 043502 (1999)
[arXiv:hep-ph/9903524]; P.~Kanti and K.~A.~Olive, ``Assisted chaotic
inflation in higher dimensional theories,'' Phys.\ Lett.\ B {\bf 464},
192 (1999) [arXiv:hep-ph/9906331].

\bibitem{shellard}
See for example: E.~P.~Shellard, A.~Vilenkin, ``Cosmic Strings and
Other Topological Defects,'' {\it Cambridge, UK: University of
Cambridge Press (1994).}, M.~B.~Hindmarsh and T.~W.~Kibble, ``Cosmic
strings,'' Rept.\ Prog.\ Phys.\ {\bf 58}, 477 (1995)
[arXiv:hep-ph/9411342].


\bibitem{wands}
D.~Wands, ``Primordial perturbations from inflation,''
arXiv:astro-ph/0201541; C.~Gordon, D.~Wands, B.~A.~Bassett and
R.~Maartens, ``Adiabatic and entropy perturbations from inflation,''
Phys.\ Rev.\ D {\bf 63}, 023506 (2001) [arXiv:astro-ph/0009131].


\bibitem{lindestuff}
 A. Linde, L. Kofman and A. Starobinsky, Prospects and Problems of
 Tachyon Cosmology {\tt hep-th/0204187} L. Kofman and A .Linde ,
 Problems with Tachyon Inflation {\tt hep-th/0205121}; M. Fairbairn
 and M. H. G. Tytgat Inflation from Tachyon Fluid ? {\tt
 hep-th/0204070}; G. Shiu and I. Wasserman, Cosmological Constraints
 on Tachyon Matter,{\tt hep-th/0205003}; G. Felder, L. Kofman and
 A. Starobinsky, Caustics in Tachyon matter and Other Born-Infeld
 Scalars, {\tt hep-th/0208019}; J Cline, H. Firouzjahl and P.
 Martineau, Reheating from Tachyon Condensation {\tt
 hep-th/0207156};C.  Kim, H. B. Kim and Y. Kim, Rolling Tachyons in
 String Cosmology {\tt hep-th/0210101 }


\bibitem{assistedgravpert}
Y.~S.~Piao, R.~G.~Cai, X.~m.~Zhang and Y.~Z.~Zhang,
``Assisted tachyonic inflation,''
Phys.\ Rev.\ D {\bf 66}, 121301 (2002)
[arXiv:hep-ph/0207143];
M.~Sami, P.~Chingangbam and T.~Qureshi,
``Aspects of tachyonic inflation with exponential potential,''
Phys.\ Rev.\ D {\bf 66}, 043530 (2002)
[arXiv:hep-th/0205179];
M.~Sami,
``Implementing power law inflation with rolling tachyon on the brane,''
arXiv:hep-th/0205146.




\bibitem{ger}
A.~A.~Gerasimov and S.~L.~Shatashvili, ``On exact tachyon
potential in open string field theory,'' JHEP {\bf 0010}, 034
(2000) [arXiv:hep-th/0009103], D.~Kutasov, M.~Marino and
G.~W.~Moore, ``Some exact results on tachyon condensation in
string field theory,'' JHEP {\bf 0010}, 045 (2000)
[arXiv:hep-th/0009148], D.~Kutasov, M.~Marino and G.~W.~Moore,
``Remarks on tachyon condensation in superstring field theory,''
arXiv:hep-th/0010108, D.~Ghoshal and A.~Sen, ``Normalisation of
the background independent open string field theory action,'' JHEP
{\bf 0011}, 021 (2000) [arXiv:hep-th/0009191]; T.~Takayanagi,
S.~Terashima and T.~Uesugi, ``Brane-antibrane action from boundary
string field theory,'' JHEP {\bf 0103}, 019 (2001)
[arXiv:hep-th/0012210];P.~Kraus and F.~Larsen, ``Boundary string
field theory of the DD-bar system,'' Phys.\ Rev.\ D {\bf 63},
106004 (2001) [arXiv:hep-th/0012198]; and for a review: K.~Ohmori,
``A review on tachyon condensation in open string field
theories,'' arXiv:hep-th/0102085.

\bibitem{zwiebachfield}
J.~A.~Minahan and B.~Zwiebach,
``Effective tachyon dynamics in superstring theory,''
JHEP {\bf 0103}, 038 (2001)
[arXiv:hep-th/0009246].

\bibitem{kogan}
S.~A.~Abel, J.~L.~Barbon, I.~I.~Kogan and E.~Rabinovici, ``String
thermodynamics in D-brane backgrounds,'' JHEP {\bf 9904}, 015
(1999) [hep-th/9902058].


\bibitem{paper4}
M.~Majumdar and A.~C.~Davis, ``D-brane Anti-brane Annihilation in
an Expanding Universe,'' [arXiv:hep-th/0304153].

\bibitem{greene}
R.~Easther, B.~R.~Greene and M.~G.~Jackson,
``Cosmological string gas on orbifolds,''
Phys.\ Rev.\ D {\bf 66}, 023502 (2002)
[arXiv:hep-th/0204099].
\end{thebibliography}
\end{document}